\documentclass[a4paper]{jpconf}

\usepackage{amsmath}
\usepackage{graphicx}
\graphicspath{{figures/}}
\usepackage{xspace}

\newcommand{\psj}[1]{\ensuremath{P_{s,#1}}}

\newcommand{\mpcinv}{\ensuremath{\, \text{Mpc}^{-1}}}
\newcommand{\esp}[1]{\times 10^{#1}}
\newcommand{\eV}{\ensuremath{\, \text{eV}}}
\newcommand{\pchip}{\texttt{PCHIP}\xspace}

\newcommand{\lcdm}{$\Lambda$CDM}

\newcommand{\DNeff}{\ensuremath{\Delta{N}_{\text{eff}}}}
\newcommand{\Neff}{\ensuremath{{N}_{\text{eff}}}}
\newcommand{\meff}[1]{\ensuremath{{m}^{\text{eff}}_{#1}}}

\newlength{\lencontours}
\begin{document}
\title{Dark Radiation and Inflationary Freedom}

\author{Stefano Gariazzo}

\address{Department of Physics, University of Torino, and INFN, Sezione di Torino, Via P. Giuria 1, I--10125 Torino, Italy}

\ead{gariazzo@to.infn.it}

\begin{abstract}
A relaxed primordial power spectrum (PPS) of scalar perturbations arising from inflation can impact
the dark radiation constraints obtained from Cosmic Microwave Background and other cosmological measurements.
If inflation produces a non-standard PPS for the initial fluctuations, 
a fully thermalized light sterile neutrino can be favoured by CMB observations, instead of being strongly disfavoured.
In the case of a thermal axion, the constraints on the axion mass are relaxed 
when the PPS is different from the standard power law.
Based on Refs.~\cite{Gariazzo:2014dla, DiValentino:2015zta}.
\end{abstract}

\section{Introduction}
In the Standard Cosmological model, 
\emph{inflation} is the initial phase of accelerated expansion.
It was introduced \cite{Guth:1980zm,Linde:1981mu}
to solve two well known problems of the Big Bang theory:
the horizon problem and the flatness problem.
Inflation occurred at energy scales that cannot be tested in laboratory, 
with the consequence that the only constraints come from cosmological observables:
among the others, the detection of a signal of primordial gravitational waves 
would be a strong evidence for the inflationary paradigm.
Several models were proposed to explain inflation, 
that is a complex process: it can have different origins and it can influence the later evolution in different ways.
Among the relics of inflation, 
the power spectrum of the initial scalar perturbations $P_s(k)$ is one of the most important,
since it gives the initial fluctuations that grow into the structures we observe today.
The simplest inflationary models predict a featureless primordial power spectrum (PPS),
that can be parameterized with a simple power-law (PL):
\begin{equation}\label{eq:plpps}
  P_s(k)=A_s \left(k/k_*\right)^{n_s-1}\,,
\end{equation}
where $n_s$ is the scalar spectral index, $A_s$ is the PPS amplitude and $k_*$ is the pivot scale.
Deviations in the inflationary process can however lead to more complicate forms for the PPS
and eventually to the presence of features and scale dependencies 
(see e.g.~Refs.~\cite{Martin:2013tda,Chluba:2015bqa, Romano:2014kla,Kitazawa:2014dya}).

The PPS of scalar perturbations can be studied only through the results of the cosmological evolution
and it cannot be probed directly:
we use mainly the Cosmic Microwave Background (CMB) radiation.
The most precise measurements we have today on the CMB spectra are from the WMAP \cite{Bennett:2012zja}
and the recent Planck \cite{Ade:2013sjv, Adam:2015rua} experiments.
The spectra obtained by both these experiments were deeply studied and 
they show a very good agreement with the well known \lcdm~theoretical model,
but there are some hints, especially in the low multipole part of the CMB spectrum,
that some unexplained deviation from the \lcdm~model can exist: 
one example is a dip in the temperature power spectrum at $20\leq\ell\leq30$, 
that is present both in the WMAP and Planck spectra.

Under the assumption of the \lcdm~model, many authors obtained constraints on the PPS 
using the CMB data from these experiments, using different techniques: for example, 
we list Refs.~\cite{Hunt:2013bha,dePutter:2014hza,Hazra:2014jwa,Ade:2015lrj}.
These works show some common results:
there is a range of wavemodes where the PPS is well in agreement with the PL form, 
but some deviations can exist below $k\simeq0.004\mpcinv$.
These features can be interpreted as deviations in the PPS arising from a non-standard inflationary model,
unless they are just statistical fluctuations.
Another possibility is that the \lcdm~model is incomplete and 
some unknown physical process generates the features in the CMB spectrum during the evolution.

We call ``inflationary freedom'' the possibility that inflation is generated in the context of non-standard mechanism
that gives possible deviations from the PL form of the PPS.
We show here that the ``inflationary freedom''
can have a strong impact on the constraints on dark radiation properties.
In particular, this contribution is based on Ref.~\cite{Gariazzo:2014dla} for the light sterile neutrino constraints, 
presented in Subsection~\ref{ssec:lsn},
and on Ref.~\cite{DiValentino:2015zta} for the thermal axion analysis, presented in Subsection~\ref{ssec:ax}.
Before discussing these results, in the next Section we describe our baseline theoretical model (Subsection~\ref{ssec:lcdm})
and our parameterization for the free PPS form (Subsection~\ref{ssec:pps}).

\section{Parameterization}

\subsection{\lcdm~model}\label{ssec:lcdm}
We base our analysis on the well established \lcdm~model, 
that is the simplest description of the observed Universe.
The \lcdm~model can be described with six parameters:
the energy density of Cold Dark Matter (CDM) $\Omega_ch^2$ and of baryons today $\Omega_bh^2$,
the reionization optical depth $\tau$,
the ratio of the sound horizon to the angular diameter distance at decoupling $\theta$.
Additionally, two parameters describe the simple power-law form for the PPS of scalar perturbations,
presented in Eq.~\ref{eq:plpps}:
its tilt $n_s$ and its amplitude $A_s$, both fixed at the pivot scale $k_*=0.05\mpcinv$.
Other quantities, such as the Hubble parameter today $H_0$, are derived from the previous ones.

When we do not mention different prescriptions, we consider the neutrino parameters as follows:
the sum of the neutrino masses $\sum m_\nu$ is fixed to the minimal value that neutrino oscillations allow 
in the case of a normal hierarchy, $\sum m_\nu=0.06$~eV, with one massive and two approximately massless neutrinos.
The effective number of relativistic species is defined with the following relation:
\begin{equation}\label{eq:neff}
  \rho_{r}=\left[1+\frac{7}{8}\left(\frac{4}{11}\right)^{4/3} \Neff\right] \rho_{\gamma}\,,
\end{equation}
where $\rho_{r}$ ($\rho_{\gamma}$) is the energy density of relativistic species (photons),
$7/8$ is for fermions, 
$(4/11)^{4/3}$ comes from the relation between the neutrino and the photon temperatures.
If the standard neutrinos are the only relativistic species in the early Universe in addition to photons
and we assume the standard thermal evolution,
the effective number of relativistic species is fixed to $\Neff^{\text{SM}}=3.046$ \cite{Mangano:2005cc}.

\subsection{Primordial Power Spectrum of Scalar Perturbations}\label{ssec:pps}
The standard inflationary models predicts a power-law (PL) form 
for the PPS of scalar and tensor modes generated in the very early Universe.
In principle, however, inflation can be more complicated and the PPS can present features:
while testing a cosmological model, if one considers the wrong PPS, 
some bias can be introduced in the results.
Testing all the possible inflationary models is beyond the scope of this work:
we rather prefer to consider a non-parametric description of the PPS 
to study how the constraints on the cosmological parameters change if the PPS is free to vary.

In order to describe a free form for the PPS of scalar perturbations, 
we use the method presented in Ref.~\cite{Gariazzo:2014dla}.
Our parameterization of the PPS is based on the choice of twelve nodes spanning a wide range of wavemodes:
we then describe the spectrum as an interpolation among them.
We use the ``piecewise cubic Hermite interpolating polynomial'' (\pchip) \cite{Fritsch:1980} 
to interpolate the twelve nodes $\psj{j}=P_s(k_j)/P_0$, where the wavemodes $k_j$ are
$\left\{\right.k_1=5\esp{-6}\mpcinv$,
$k_2=10^{-3} \mpcinv$,
$k_j     = k_2 (k_{11}/k_2)^{(j-2)/9}$ for $j\in[3,10]$,
$k_{11}  = 0.35 \mpcinv$,
$k_{12}  = 10\mpcinv\left.\right\}$.
In the range $(\log k_2,\, \log k_{11})$ the nodes are equally spaced, 
since this is the wavemode range were the data constraints are stronger \cite{dePutter:2014hza}.
We fix the first and last nodes to ensure that all the evaluations of the PPS are inside the covered range:
consequently, we expect that $\psj{1}$ and $\psj{12}$ are less constrained by the data.

The interpolating function we consider 
is a modified version of the \pchip~algorithm \cite{Fritsch:1984} that have the aim
to maintain in the interpolated function the same monotonicity of the initial point series.
The \pchip~PPS is:
\begin{equation}
P_{s}(k)=P_0\times\pchip(k; \psj{1}, \ldots, \psj{12})
, \label{eq:PPS_pchip}
\end{equation}
where $\psj{j}$ is the value of the PPS at the node $k_j$ divided by the normalization $P_0$.
For a complete description of the \pchip~parameterization, we refer to Ref.~\cite{Gariazzo:2014dla}.

When we consider the ``inflationary freedom'' model, with the PPS described by the \pchip~algorithm,
we use four out of six of the \lcdm~parameters described in the previous subsection 
($\Omega_ch^2$, $\Omega_bh^2$, $\tau$, $\theta$),
while the two parameters describing the PL PPS are replaced by the twelve nodes ($\psj{1},\,\ldots,\,\psj{12}$)
that we use to describe the \pchip~PPS \cite{Gariazzo:2014dla, DiValentino:2015zta}.

\section{Results}
In this section we present the results obtained for two different candidates contributing to dark radiation:
the light sterile neutrino in Subsection~\ref{ssec:lsn} and the thermal axion in Subsection~\ref{ssec:ax}.

\subsection{Light Sterile Neutrino Analysis}\label{ssec:lsn}
\looseness=-1
The first dark radiation candidate we consider is the light sterile neutrino.
This is motivated by the anomalies registered in different Short BaseLine (SBL) oscillation experiments,
such as LNSD \cite{Aguilar:2001ty} and MiniBoone \cite{Aguilar-Arevalo:2013pmq}, 
but also in the measured fluxes coming from many nuclear reactors \cite{Mention:2011rk} and 
in the calibration of some Gallium experiments for solar neutrino oscillations \cite{Giunti:2010zu}:
all these anomalies can be explained if the standard three neutrino paradigm is extended with 
the addition of a fourth neutrino mass eigenstate, providing a squared mass difference 
$\Delta m^2_{41}=\Delta m^2_{\rm{SBL}}\simeq1\eV$%
\footnote{We name $m_i$ the mass of the $i$-th neutrino mass eigenstate ($i=1,\ldots,4$) and
we define $\Delta m^2_{ij}=m^2_j-m^2_i$.}
with respect to the lightest mass eigenstate.
The fourth neutrino is mainly sterile. 
A sterile neutrino is not coupled to the Standard Model with the standard weak interactions.
For a detailed discussion on the light sterile neutrino and its effects in cosmology,
see the recent review \cite{Gariazzo:2015rra}.

To study the light sterile neutrino in cosmology, we consider an extension of the \lcdm~model, where
we add a neutrino state $\nu_s$ described by two additional parameters:
its mass $m_s$ and its contribution to the radiation energy density in the early Universe $\DNeff$.
We assume that the masses of the three lighter neutrinos are much below the mass of the additional neutrino:
in this case we can approximate 
$m_1\simeq0$ and $m_s\simeq m_4\simeq \sqrt{\Delta m^2_{\rm{SBL}}}$.
Assuming that the light sterile neutrino is the only relativistic species 
beyond the three active neutrinos and photons,
its contribution to radiation energy density is $\DNeff=\Neff-\Neff^{\text{SM}}$.
In this analysis $\DNeff$ is limited in the range $[0,1]$, since we assume
that the contribution to the radiation energy density from the additional sterile neutrino 
cannot overcome the contribution of a single standard active neutrino.

When we study the additional neutrino, we shall include information on its mass $m_s$ coming 
from the analysis of the SBL data as presented in \cite{Giunti:2013aea}:
this information is added as a prior on $m_s$ in the cosmological analysis
\cite{Archidiacono:2012ri}.

In this analysis we considered the
CMB temperature data from the Planck 2013 release \cite{Ade:2013kta}, 
the Atacama Cosmology Telescope (ACT) \cite{Dunkley:2013vu}
and the South Pole Telescope (SPT) \cite{Story:2012wx},
plus the CMB polarization data measured by WMAP \cite{Bennett:2012zja}.
We include also measurements at low redshift:
the matter power spectrum from the WiggleZ Dark Energy Survey \cite{Parkinson:2012vd},
a prior on the Hubble parameter $H_0 =73.8\pm2.4 \, \text{km} \, \text{s}^{-1} \, \text{Mpc}^{-1}$ \cite{Riess:2011yx},
the cluster mass function obtained by the Planck cluster counts through the Sunyaev Zel'dovich (SZ) effect \cite{Ade:2013lmv}
and the CFHTLenS measurements of the 2D cosmic shear \cite{Kilbinger:2012qz,Heymans:2013fya}.
This complete dataset is named ``COSMO''.

\begin{figure}
  \begin{center}
  \includegraphics[width=\lencontours]{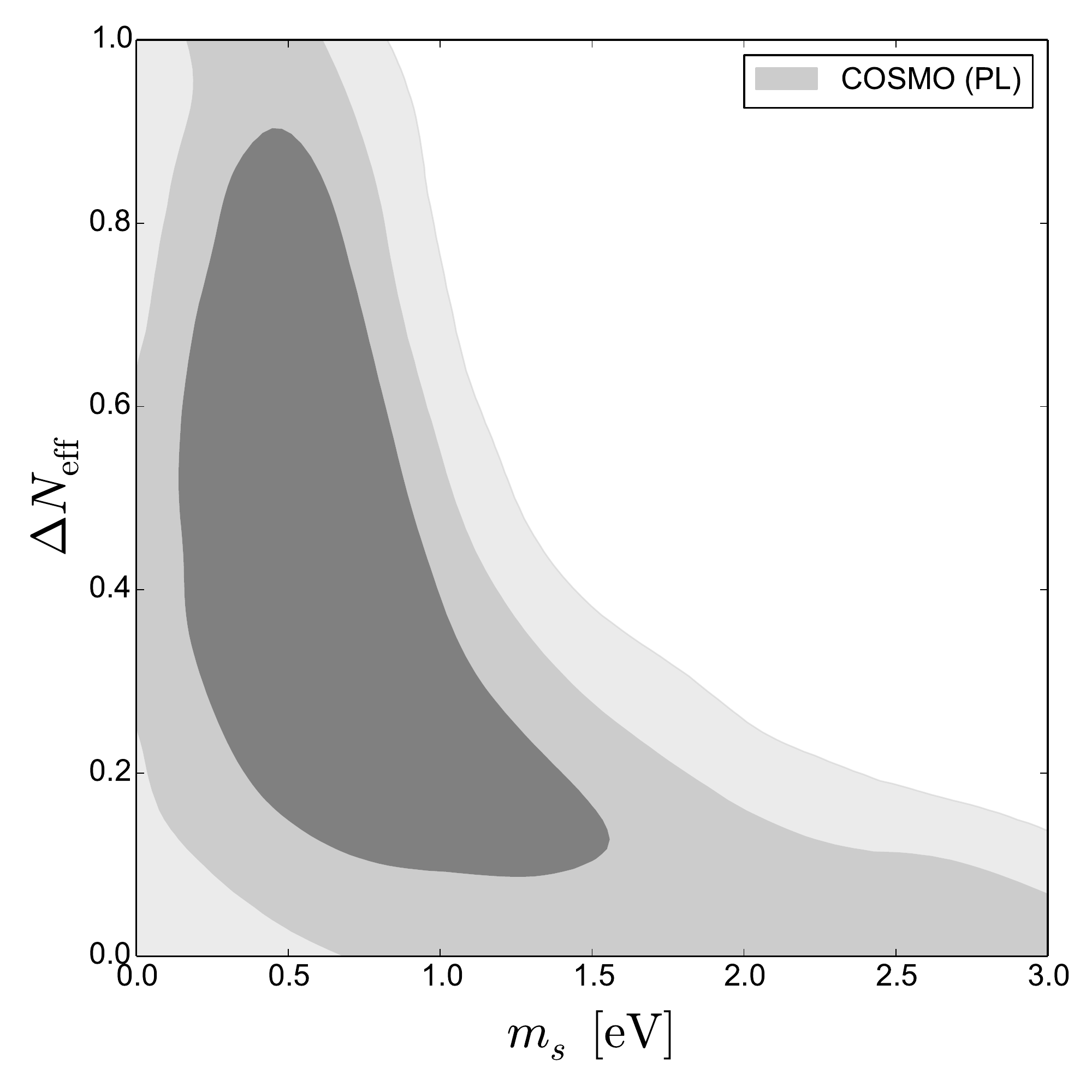}
  \includegraphics[width=\lencontours]{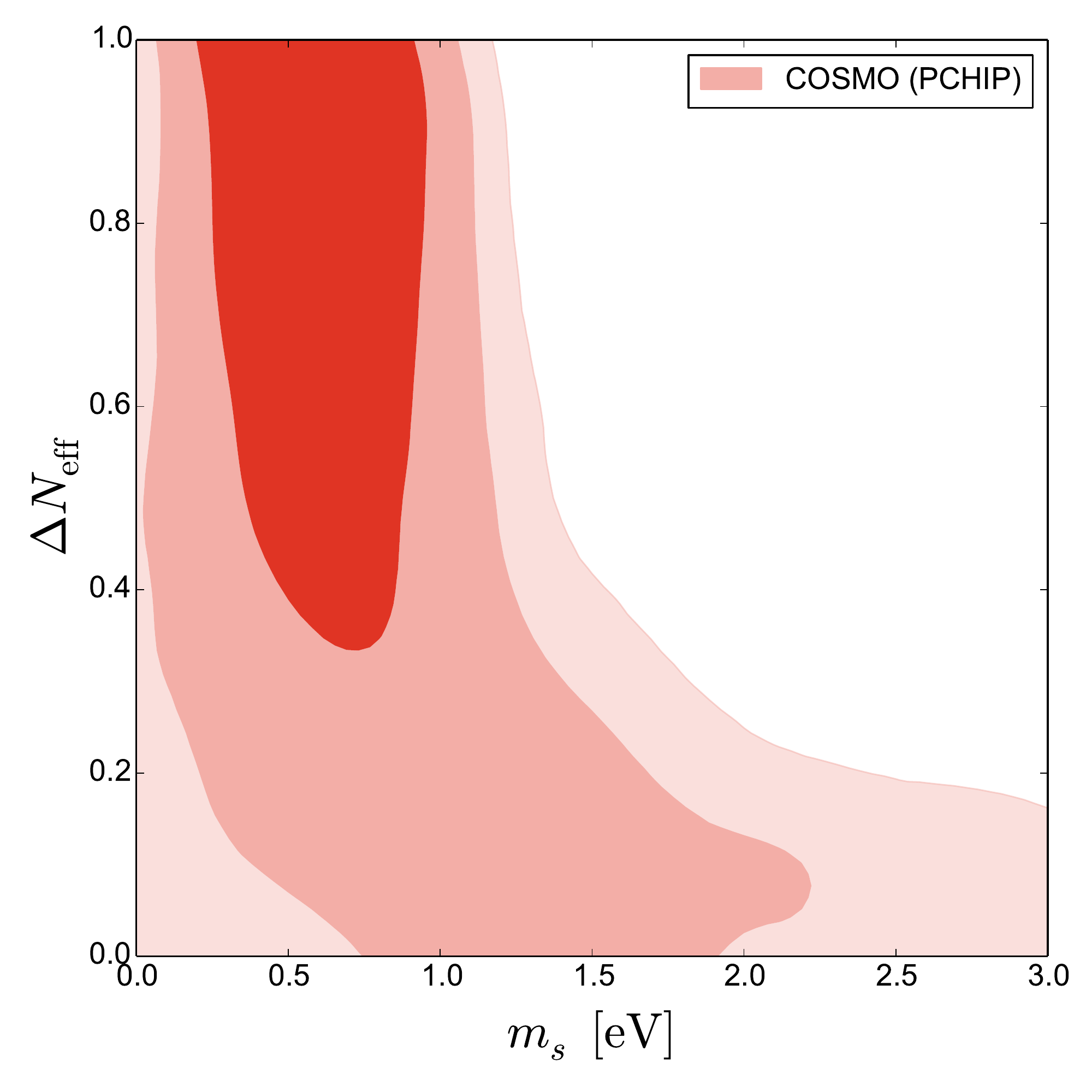}
  \end{center}
  \caption{Marginalized constraints at 1, 2 and 3$\sigma$ 
  in the (\meff{s}, \DNeff) plane,
  from cosmological data only.
  In the left panel we considered the standard power-law (PL) form for the PPS of scalar perturbations,
  while in the right panel this assumption is relaxed and we used a \pchip~PPS.
  From~\cite{Gariazzo:2014dla}.}
  \label{fig:nu_nosbl}
\end{figure}

In Figure~\ref{fig:nu_nosbl} we compare the marginalized constraints in the ($m_s$, $\Neff$) plane
obtained in the context of the extended \lcdm+$\nu_s$~model
when a standard PL PPS is assumed (left panel) with those obtained with the free \pchip~PPS (right panel).
The preference for $m_s>0$ is driven by the inclusion of the cluster counts and cosmic shear measurement,
while the preference for $\DNeff>0$ is driven by the correlation with the Hubble parameter $H_0$ and 
the inclusion of the $H_0$ prior (see e.g.~Refs.~\cite{Gariazzo:2013gua,Archidiacono:2014apa}).
Nonetheless, $\DNeff=1$ is disfavoured when the PPS is described by the PL form,
with the consequence that a full thermalization with the active neutrinos is disfavoured for the sterile neutrino.
In the right panel we can compare how the results change when a \pchip~PPS is included:
the additional freedom in the PPS can compensate the Silk damping effect driven by the higher $\Neff$
and the favoured value is $\DNeff=1$, corresponding to a fully thermalized sterile neutrino.

\begin{figure}
  \begin{center}
  \includegraphics[width=\lencontours]{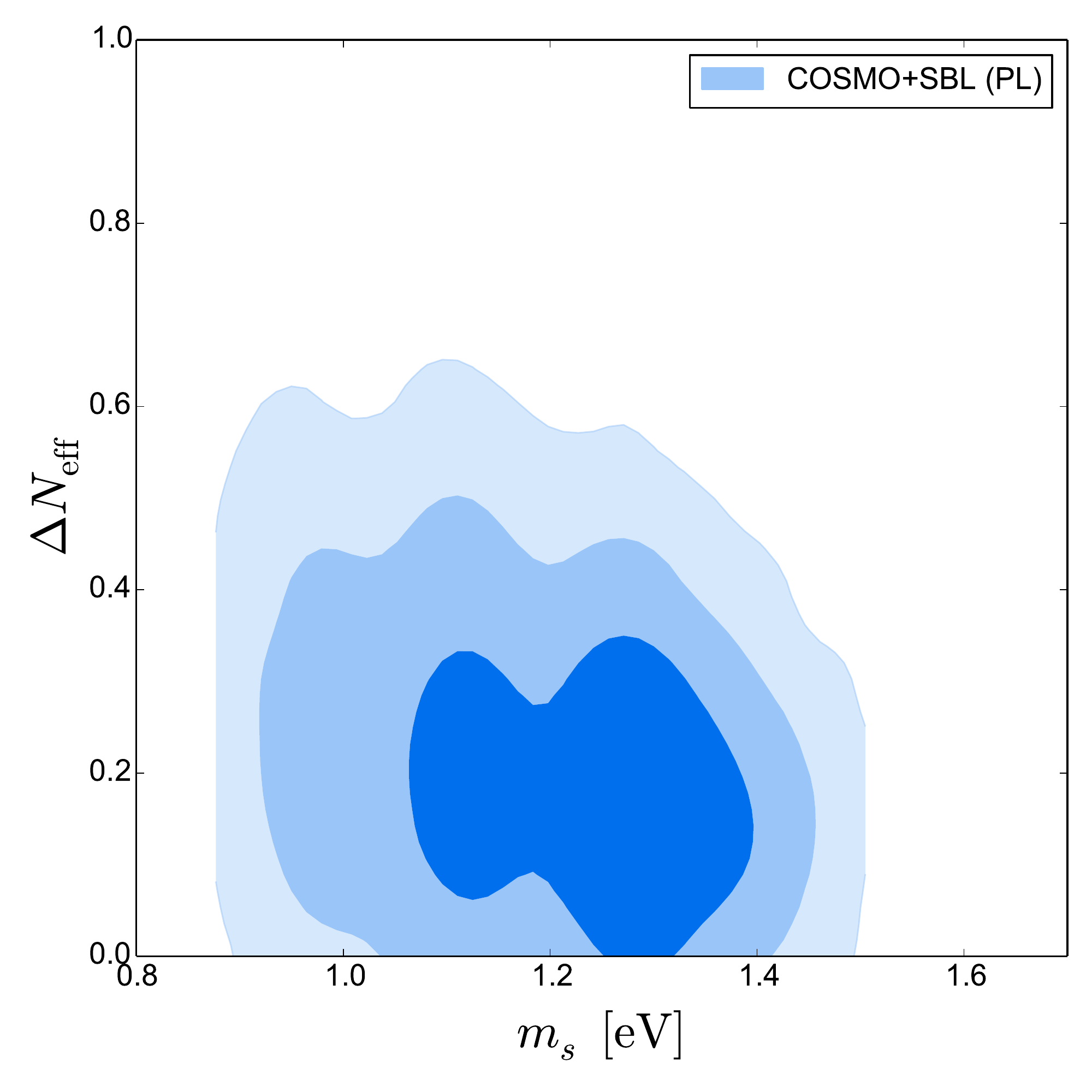}
  \includegraphics[width=\lencontours]{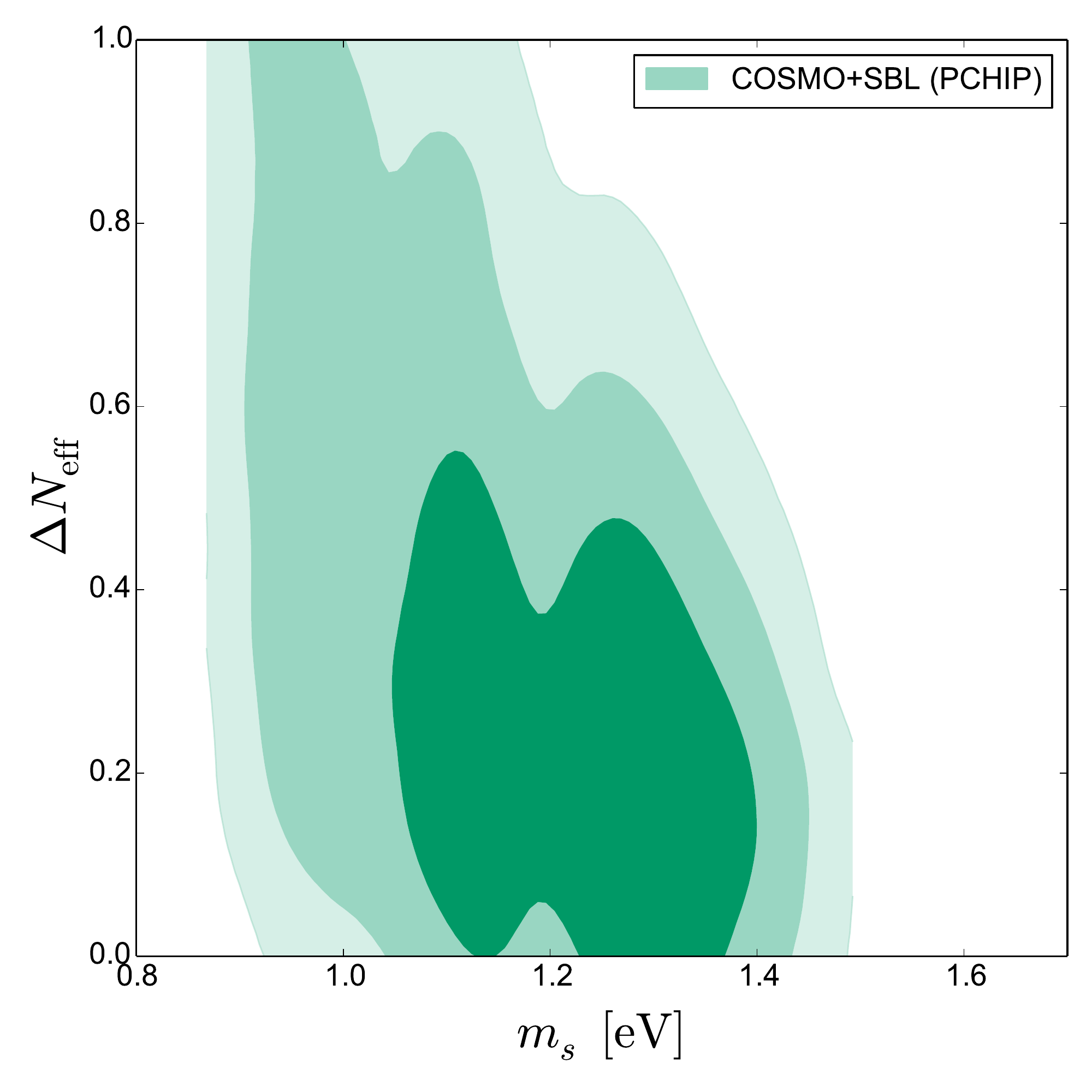}
  \end{center}
  \caption{As in Fig.~\ref{fig:nu_nosbl}, 
  but with the inclusion of the prior on $m_s$ from SBL experiments \cite{Giunti:2013aea}.
  From~\cite{Gariazzo:2014dla}.}
  \label{fig:nu_sbl}
\end{figure}

The situation is slightly different for the case in Fig.~\ref{fig:nu_sbl}, 
where the only difference with Fig.~\ref{fig:nu_nosbl}
is that the fit is performed using the SBL prior on $m_s$ in addition to the COSMO dataset.
In the left panel, obtained with the PL PPS, 
we see how for a $m_s\simeq1\eV$ neutrino $\DNeff=1$ is strongly disfavoured,
with an upper limit of roughly $\DNeff<0.5$ at 3$\sigma$.
Also in this case the introduction of inflationary freedom with the \pchip~PPS changes the situation:
even if $\DNeff=1$ is still disfavoured by the 1$\sigma$ constraints, the full thermalization of the sterile neutrino is
compatible with the 2$\sigma$ constraints, if $m_s$ is slightly less than 1~eV.
If the sterile neutrino existence will be confirmed by the SBL oscillation experiments,
we will have to deal with its presence in cosmology:
a non-standard shape of the PPS would possibly reconcile a full thermalization with cosmology,
solving the theoretical problems 
related to the explanation of the sterile neutrino incomplete thermalization.

\subsection{Axion sector}\label{ssec:ax}
The second candidate of dark radiation we consider here is the thermal axion.
Axions represent a possible solution to the strong CP problem \cite{Peccei:1977hh, Peccei:1977ur}, 
being the Pseudo-Nambu-Goldstone bosons of a global symmetry $U(1)_{PQ}$ (Peccei-Quinn),
spontaneously broken at the axion scale~$f_a$.
Axions can be produced either in non-thermal or thermal processes:
here we consider only the latter possibility.
A thermally produced axion has an impact on cosmology similar to a extra sterile neutrino,
contributing as radiation in the early Universe and as a massive component at late times.
As for the sterile neutrino, if the thermal axion is the only relativistic species beyond photons and active neutrinos,
its contribution to the radiation energy density is $\DNeff=\Neff-3.046$:
this quantity can be calculated numerically in cosmology and depends on $f_a$.
The mass of the axion depends on the scale $f_a$ through the relation 
\begin{equation}\label{eq:axmass}
m_a = 
0.6\ {\rm eV}\ \frac{10^7\, {\rm GeV}}{f_a}~
.
\end{equation}
Since both the axion mass and the axion effective number $\DNeff$ can be expressed as a function of the axion scale $f_a$,
for practical reasons, we parameterize $\DNeff$ as a function of $m_a$ and we use only the axion mass 
as an additional parameter to extend the \lcdm~model.
For a detailed calculation of the axion properties in cosmology, see e.g.~Ref.~\cite{Hannestad:2005df}.
For recent constraints on the axion mass from cosmology, see e.g.~Ref.~\cite{DiValentino:2015wba}.

In this analysis we considered as baseline dataset
the same CMB data mentioned for the sterile neutrino analysis:
temperature spectra from the Planck 2013 release \cite{Ade:2013kta}, 
the Atacama Cosmology Telescope (ACT) \cite{Dunkley:2013vu}
and the South Pole Telescope (SPT) \cite{Story:2012wx},
plus the CMB polarization data measured by WMAP \cite{Bennett:2012zja}.
We shall extend it with 
a prior on the Hubble parameter $H_0 =70.6\pm3.3 \, \text{km} \, \text{s}^{-1} \, \text{Mpc}^{-1}$ \cite{Efstathiou:2013via},
and Baryon Acoustic Oscillations (BAO) data from 
the WiggleZ~\cite{Blake:2011en}, the
6dF~\cite{Beutler:2011hx} and the SDSS II surveys~\cite{Percival:2009xn,Padmanabhan:2012hf},
plus the DR11 results from the
Baryon Oscillation Spectroscopic Survey (BOSS)~\cite{Anderson:2013zyy} survey,
and using the results from the Planck cluster counts (PSZ) \cite{Ade:2013lmv}, 
obtained both with a fixed or a free mass bias.

\begin{figure}
  \begin{center}
  \includegraphics[width=\lencontours]{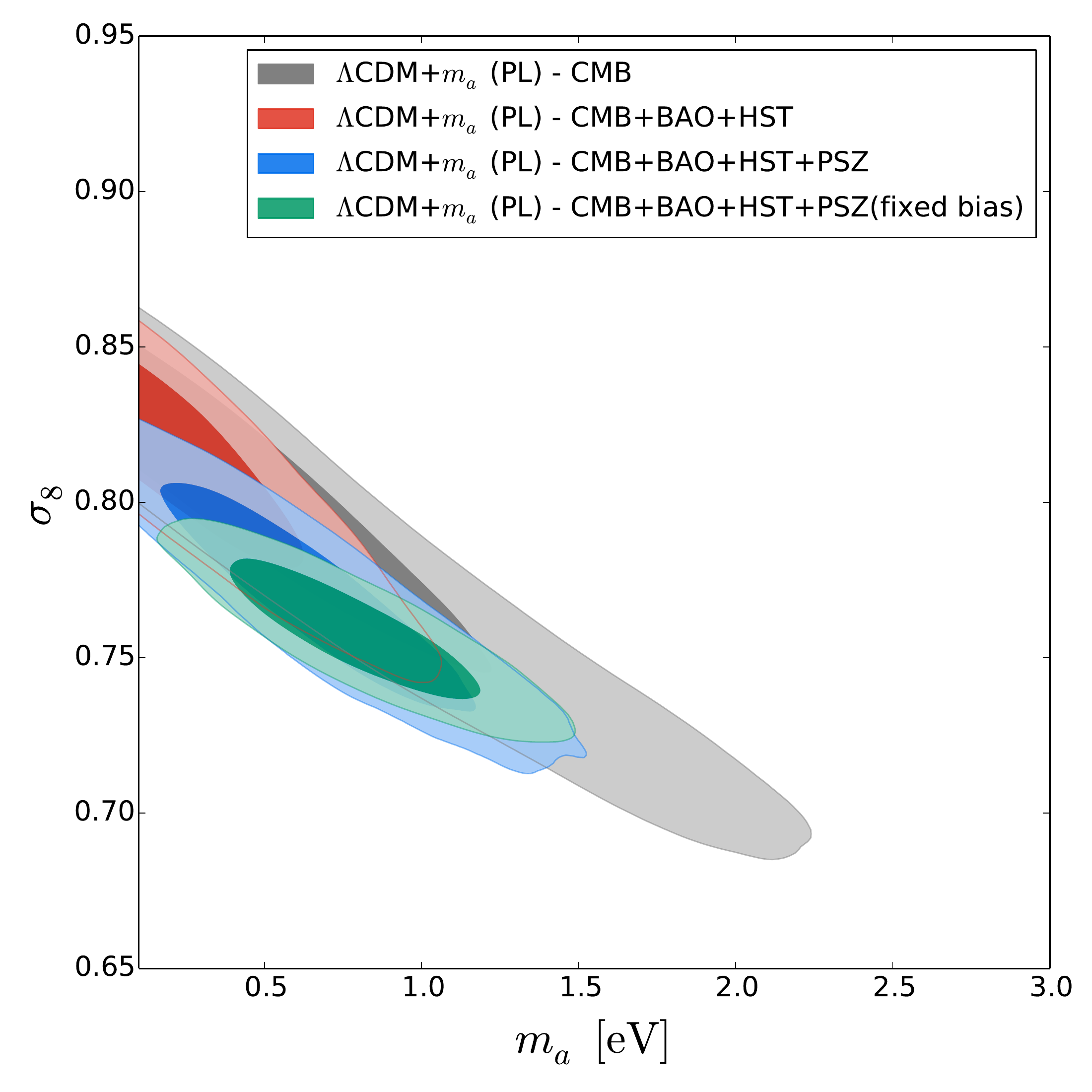}
  \includegraphics[width=\lencontours]{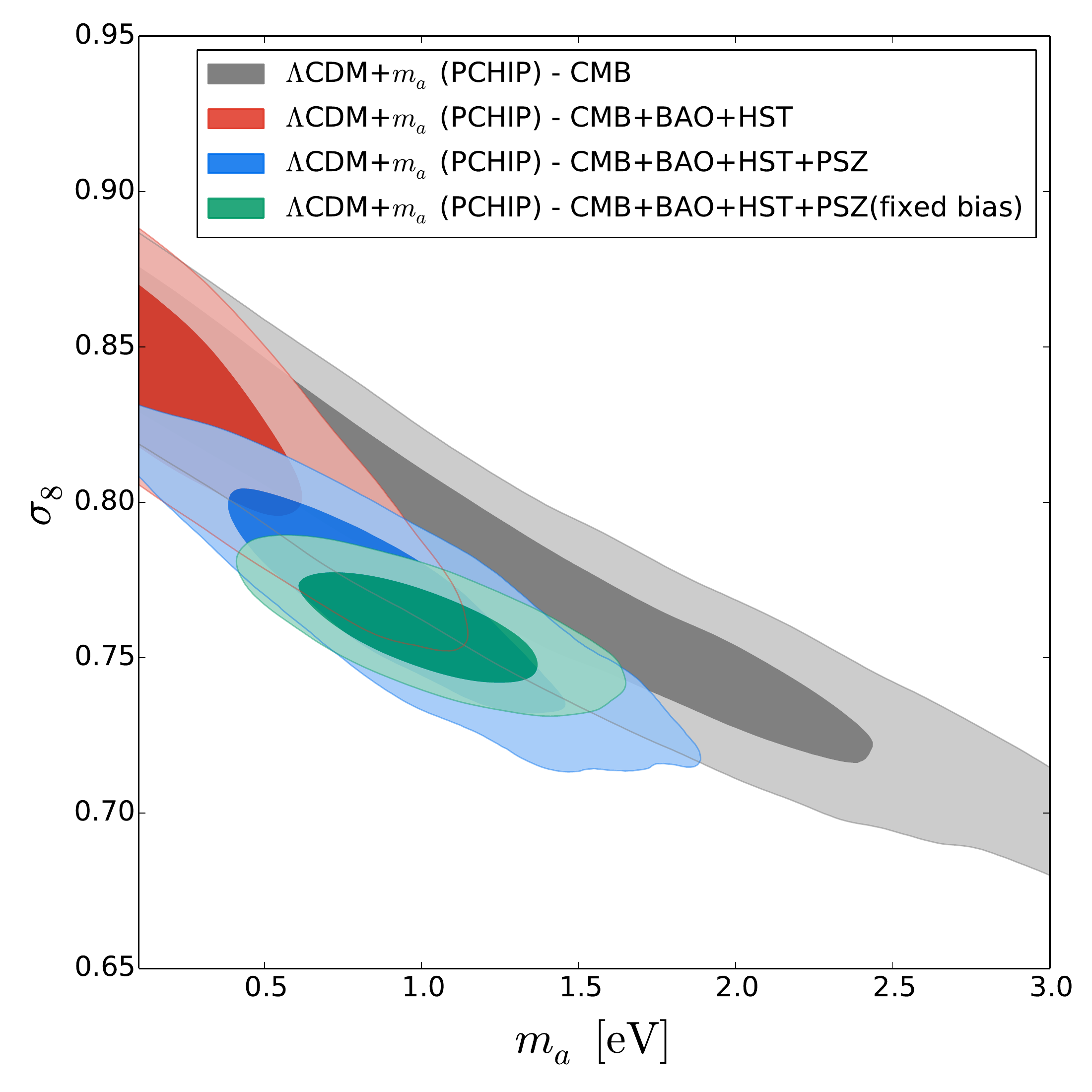}
  \end{center}
  \caption{Marginalized constraints at 1 and 2$\sigma$ 
  in the ($m_a$, $\sigma_8$) plane,
  from different cosmological data combinations.
  In the left panel we considered the standard power-law (PL) form for the PPS of scalar perturbations,
  while in the right panel this assumption is relaxed and we used a \pchip~PPS.
  From~\cite{DiValentino:2015zta}.}
  \label{fig:ax_bounds}
\end{figure}

We show in Fig.~\ref{fig:ax_bounds} the marginalized constraints at 1 and 2$\sigma$ in the ($m_a$, $\sigma_8$) plane,
for different data combinations, when using the PL PPS (left panel) and the \pchip~PPS (right panel).
Similarly to what we found in the neutrino analysis, the axion mass constraints are relaxed when the PPS is free,
but in this case the difference is not so large.
The main difference is that in this case the axion contribution to $\Neff$ depends on $m_a$, 
while for the sterile neutrino the mass contribution is independent of $\DNeff$:
giving rise to two well separated effects, the axion mass constraints are more robust and 
less influenced by inflationary freedom.
The freedom in the PPS, in fact, can compensate in an easier way the Silk damping that comes from an increase of $\Neff$
than the effects driven by the addition of a massive component.

\subsection{Constraints on the Free Primordial Power Spectrum}
\begin{figure}
  \begin{center}
  \includegraphics[width=0.49\textwidth]{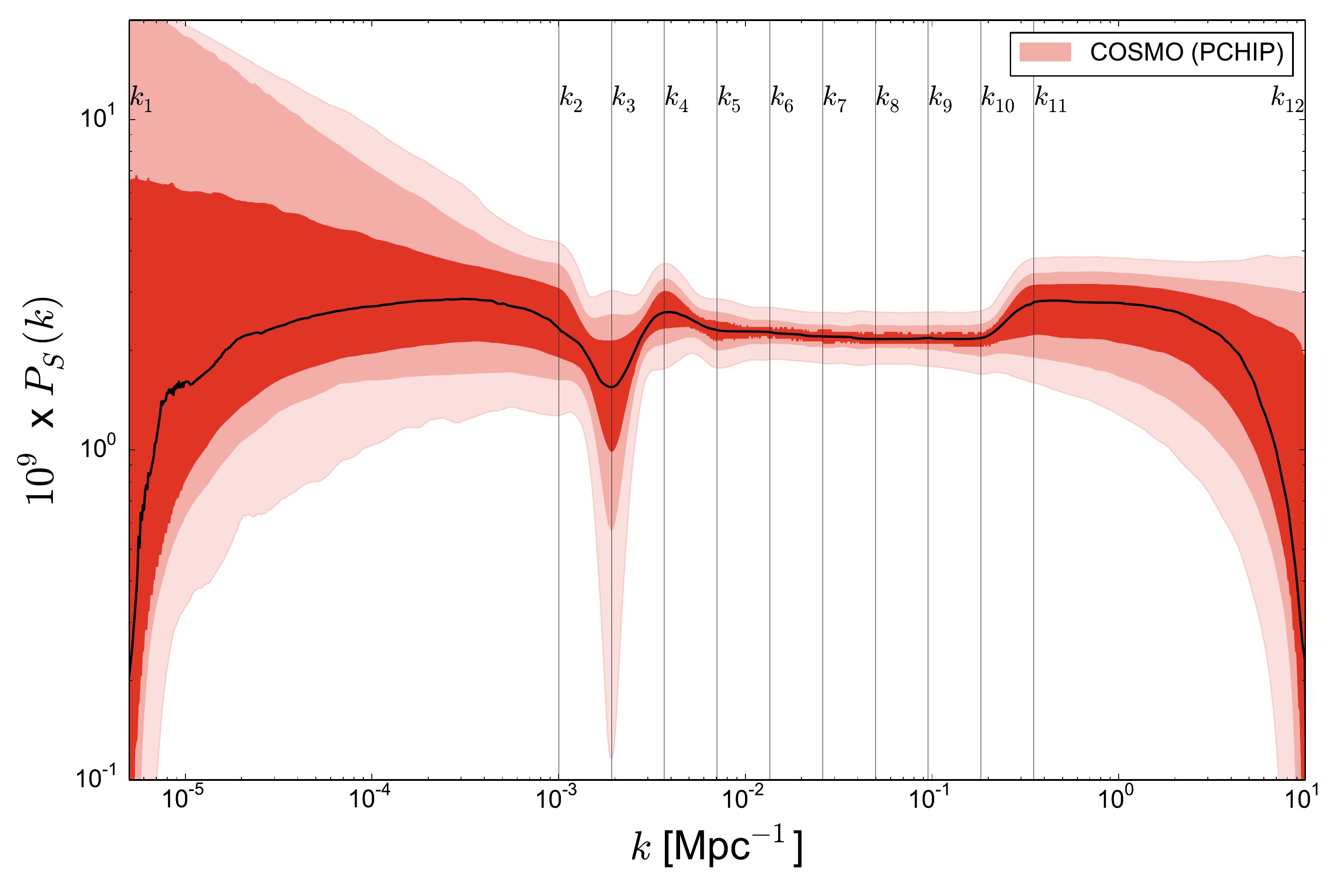}
  \includegraphics[width=0.49\textwidth]{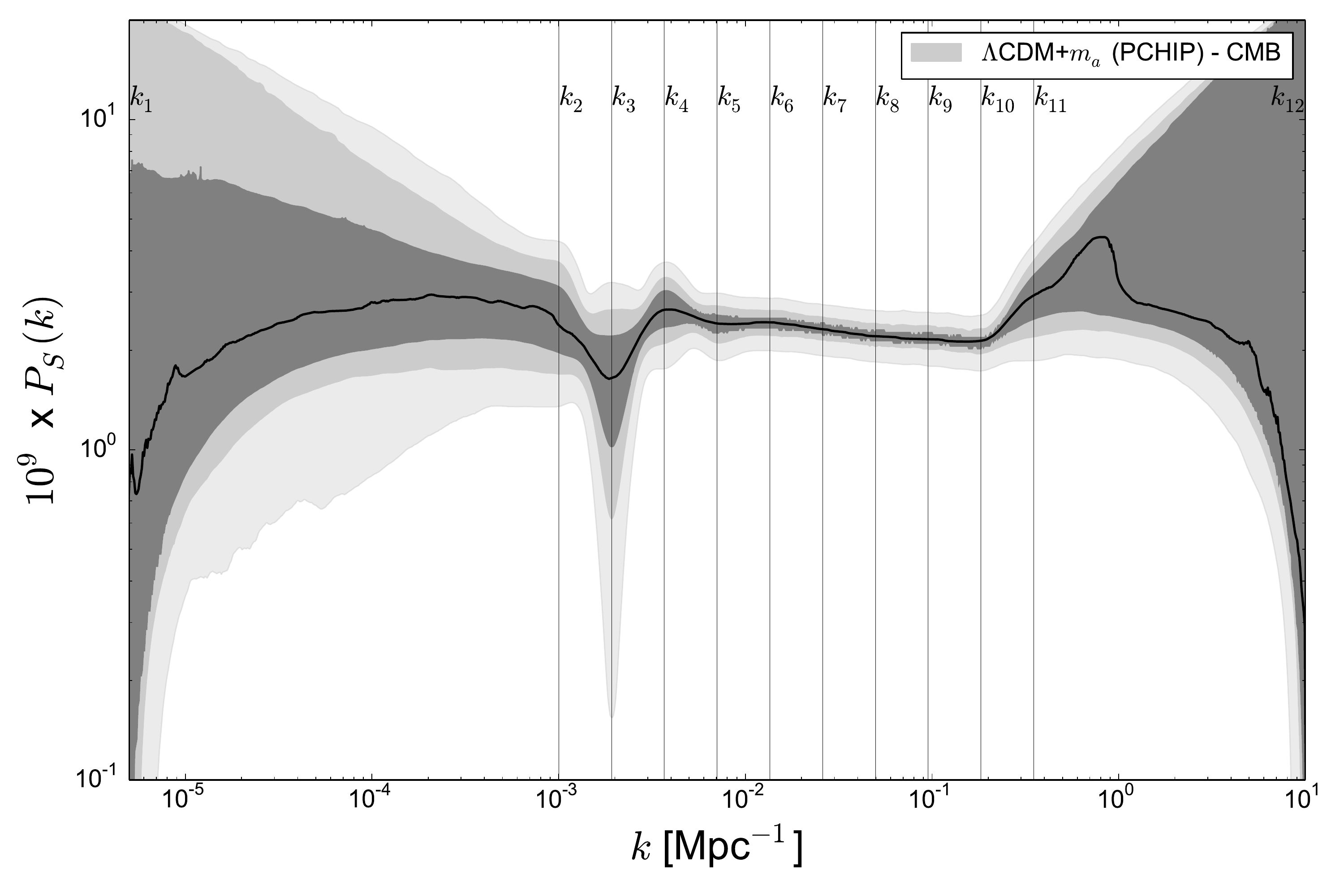}
  \end{center}
  \caption{Shape of the reconstructed \pchip~PPS,
  in the \lcdm+$\nu_s$ model (left panel, from \cite{Gariazzo:2014dla}) and 
  in the \lcdm+$m_a$ model (right panel, from \cite{DiValentino:2015zta}).
  The solid line represents the best-fit, while the bands are the contours at 1, 2, 3$\sigma$.
  }
  \label{fig:pps}
\end{figure}

From the MCMC analyses shown in the previous sections
we obtain also constraints on the shape of the \pchip~PPS.
In Fig.~\ref{fig:pps} we compare the \pchip~PPS reconstructions as obtained in the
\lcdm+$\nu_s$ model (left panel, from \cite{Gariazzo:2014dla}) and 
in the \lcdm+$m_a$ model (right panel, from \cite{DiValentino:2015zta}).
We can list a number of common properties for these two spectra:
there is a region that is well approximated by the PL~PPS, 
that is the part included in the range $7\esp{-3}\mpcinv\leq k\leq 0.2\mpcinv$:
the difference between the two reconstructions, here, is the tilt of the PPS, 
that must be closer to be flat in the \lcdm+$\nu_s$ 
in order to compensate a higher $\DNeff$ from the sterile neutrino.
Outside this range the PPS contains a number of features:
the main deviations from the PL form are located around 
$k=2\esp{-3}\mpcinv$, where a dip corresponding to the one at $\ell\simeq22$ in the CMB spectrum is present,
and $k=3.5\esp{-3}\mpcinv$, with a small bump corresponing to the one 
in the CMB spectrum at $\ell\simeq40$.
The behaviour at $k\leq10^{-4}\mpcinv$ and $k\geq1\mpcinv$ is more uncertain 
since it is poorly constrained by the cosmological data.
These results are in agreement with the constraints obtained 
with different and possibly more precise reconstruction methods
(see e.g.~Refs.~\cite{Hunt:2013bha,dePutter:2014hza,Hazra:2014jwa,Ade:2015lrj}).

\section{Conclusions}
Inflation is the mechanism that provides a solution to the flatness and horizon problems 
with an initial phase of accelerated expansion in the early Universe.
Even if the simplest inflationary models predict a featureless primordial power spectrum (PPS) for scalar perturbations,
Nature can have chosen a more complicate model, with the consequence that the PPS is not featureless and 
it can only be approximated by a power-law description in a limited range of wavemodes.
If this is the case, the freedom in the PPS can have a strong impact
on the constraints of other cosmological quantities,
such as the properties of dark radiation.
We showed how the constraints on a light sterile neutrino or a thermal axion can change:
if the PPS is free, a fully thermalized light sterile neutrino is preferred from the cosmological analysis, 
instead of being strongly disfavoured as when the PPS is described by the power-law.
The constraints on the thermal axion mass are also relaxed when the PPS is relaxed from the power-law to the \pchip~form,
even if in a less significant way.
A new analysis of dark radiation properties with the inclusion of the newest 2015 Planck results 
is currently in preparation \cite{diValentino:2015inprep}.

We considered here how ``inflationary freedom'' can change the constraints on dark radiation properties,
but it can be interesting to extend the study to different cosmological observables.
As an example, a non-standard PPS can influence the constraints on non-gaussianities from future probes \cite{Gariazzo:2015qea}:
the missing knowledge about inflation, then, can lead to possible biases 
in the constraints for a number of cosmological quantities.

\section*{Acknowledgements}
The work of S. G. was supported by the Theoretical Astroparticle Physics research
Grant No. 2012CPPYP7 under the Program PRIN 2012 funded by the Ministero
dell'Istruzione, Universit\`a e della Ricerca (MIUR).

\section*{References}
\bibliographystyle{iopart-num}
\bibliography{taup}

\providecommand{\newblock}{}
\begin{thebibliography}{10}
\expandafter\ifx\csname url\endcsname\relax
  \def\url#1{{\tt #1}}\fi
\expandafter\ifx\csname urlprefix\endcsname\relax\def\urlprefix{URL }\fi
\providecommand{\eprint}[2][]{\url{#2}}

\bibitem{Gariazzo:2014dla}
Gariazzo S, Giunti C and Laveder M 2015 {\em JCAP\/} {\bf 1504} 023
  (\textit{Preprint} \eprint{1412.7405})

\bibitem{DiValentino:2015zta}
Di~Valentino E, Gariazzo S, Giusarma E and Mena O 2015 {\em Phys. Rev.\/} {\bf
  D91} 123505 (\textit{Preprint} \eprint{1503.00911})

\bibitem{Guth:1980zm}
Guth A~H 1981 {\em Phys. Rev.\/} {\bf D23} 347--356

\bibitem{Linde:1981mu}
Linde A~D 1982 {\em Phys. Lett.\/} {\bf B108} 389--393

\bibitem{Martin:2013tda}
Martin J, Ringeval C and Vennin V 2014 {\em Phys. Dark Univ.\/} {\bf 5-6}
  75--235 (\textit{Preprint} \eprint{1303.3787})

\bibitem{Chluba:2015bqa}
Chluba J, Hamann J and Patil S~P 2015 {\em Int. J. Mod. Phys.\/} {\bf D24}
  1530023 (\textit{Preprint} \eprint{1505.01834})

\bibitem{Romano:2014kla}
Romano A~E and Cadavid A~G 2014  (\textit{Preprint} \eprint{1404.2985})

\bibitem{Kitazawa:2014dya}
Kitazawa N and Sagnotti A 2014 {\em JCAP\/} {\bf 1404} 017 (\textit{Preprint}
  \eprint{1402.1418})

\bibitem{Bennett:2012zja}
Bennett C~L {\em et~al.\/} (WMAP) 2013 {\em Astrophys. J. Suppl.\/} {\bf 208}
  20 (\textit{Preprint} \eprint{1212.5225})

\bibitem{Ade:2013sjv}
Ade P~A~R {\em et~al.\/} 2014 {\em Astron. Astrophys.\/} {\bf 571} A1
  (\textit{Preprint} \eprint{1303.5062})

\bibitem{Adam:2015rua}
Adam R {\em et~al.\/} 2015  (\textit{Preprint} \eprint{1502.01582})

\bibitem{Hunt:2013bha}
Hunt P and Sarkar S 2014 {\em JCAP\/} {\bf 1401} 025 (\textit{Preprint}
  \eprint{1308.2317})

\bibitem{dePutter:2014hza}
de~Putter R, Linder E~V and Mishra A 2014 {\em Phys.Rev.\/} {\bf D89} 103502
  (\textit{Preprint} \eprint{1401.7022})

\bibitem{Hazra:2014jwa}
Hazra D~K, Shafieloo A and Souradeep T 2014 {\em JCAP\/} {\bf 1411} 011
  (\textit{Preprint} \eprint{1406.4827})

\bibitem{Ade:2015lrj}
Ade P~A~R {\em et~al.\/} (Planck Collaboration) 2015  (\textit{Preprint}
  \eprint{1502.02114})

\bibitem{Mangano:2005cc}
Mangano G {\em et~al.\/} 2005 {\em Nucl. Phys.\/} {\bf B729} 221--234
  (\textit{Preprint} \eprint{hep-ph/0506164})

\bibitem{Fritsch:1980}
Fritsch F and Carlson R 1980 {\em SIAM Journal on Numerical Analysis\/} {\bf
  17} 238

\bibitem{Fritsch:1984}
Fred~Fritsch J~B 1984 {\em SIAM Journal on Scientific and Statistical
  Computing\/} {\bf 5} 300

\bibitem{Aguilar:2001ty}
Aguilar-Arevalo A {\em et~al.\/} (LSND) 2001 {\em Phys. Rev.\/} {\bf D64}
  112007 (\textit{Preprint} \eprint{hep-ex/0104049})

\bibitem{Aguilar-Arevalo:2013pmq}
Aguilar-Arevalo A~A {\em et~al.\/} (MiniBooNE) 2013 {\em Phys. Rev. Lett.\/}
  {\bf 110} 161801 (\textit{Preprint} \eprint{1207.4809})

\bibitem{Mention:2011rk}
Mention G {\em et~al.\/} 2011 {\em Phys. Rev.\/} {\bf D83} 073006
  (\textit{Preprint} \eprint{1101.2755})

\bibitem{Giunti:2010zu}
Giunti C and Laveder M 2011 {\em Phys. Rev.\/} {\bf C83} 065504
  (\textit{Preprint} \eprint{1006.3244})

\bibitem{Gariazzo:2015rra}
Gariazzo S, Giunti C, Laveder M, Li Y~F and Zavanin E~M 2015
  (\textit{Preprint} \eprint{1507.08204})

\bibitem{Giunti:2013aea}
Giunti C, Laveder M, Li Y~F and Long H~W 2013 {\em Phys. Rev.\/} {\bf D88}
  073008 (\textit{Preprint} \eprint{1308.5288})

\bibitem{Archidiacono:2012ri}
Archidiacono M, Fornengo N, Giunti C and Melchiorri A 2012 {\em Phys. Rev.\/}
  {\bf D86} 065028 (\textit{Preprint} \eprint{1207.6515})

\bibitem{Ade:2013kta}
Ade P~A~R {\em et~al.\/} 2014 {\em Astron. Astrophys.\/} {\bf 571} A15
  (\textit{Preprint} \eprint{1303.5075})

\bibitem{Dunkley:2013vu}
Dunkley J {\em et~al.\/} 2013 {\em JCAP\/} {\bf 1307} 025 (\textit{Preprint}
  \eprint{1301.0776})

\bibitem{Story:2012wx}
Story K~T {\em et~al.\/} 2013 {\em Astrophys. J.\/} {\bf 779} 86
  (\textit{Preprint} \eprint{1210.7231})

\bibitem{Parkinson:2012vd}
Parkinson D {\em et~al.\/} 2012 {\em Phys. Rev.\/} {\bf D86} 103518
  (\textit{Preprint} \eprint{1210.2130})

\bibitem{Riess:2011yx}
Riess A~G {\em et~al.\/} 2011 {\em Astrophys. J.\/} {\bf 730} 119
  (\textit{Preprint} \eprint{1103.2976})

\bibitem{Ade:2013lmv}
Ade P~A~R {\em et~al.\/} (Planck) 2014 {\em Astron. Astrophys.\/} {\bf 571} A20
  (\textit{Preprint} \eprint{1303.5080})

\bibitem{Kilbinger:2012qz}
Kilbinger M {\em et~al.\/} 2013 {\em Mon. Not. Roy. Astron. Soc.\/} {\bf 430}
  2200--2220 (\textit{Preprint} \eprint{1212.3338})

\bibitem{Heymans:2013fya}
Heymans C {\em et~al.\/} 2013 {\em Mon. Not. Roy. Astron. Soc.\/} {\bf 432}
  2433 (\textit{Preprint} \eprint{1303.1808})

\bibitem{Gariazzo:2013gua}
Gariazzo S, Giunti C and Laveder M 2013 {\em JHEP\/} {\bf 11} 211
  (\textit{Preprint} \eprint{1309.3192})

\bibitem{Archidiacono:2014apa}
Archidiacono M {\em et~al.\/} 2014 {\em JCAP\/} {\bf 1406} 031
  (\textit{Preprint} \eprint{1404.1794})

\bibitem{Peccei:1977hh}
Peccei R~D and Quinn H~R 1977 {\em Phys. Rev. Lett.\/} {\bf 38} 1440--1443

\bibitem{Peccei:1977ur}
Peccei R~D and Quinn H~R 1977 {\em Phys. Rev.\/} {\bf D16} 1791--1797

\bibitem{Hannestad:2005df}
Hannestad S, Mirizzi A and Raffelt G 2005 {\em JCAP\/} {\bf 0507} 002
  (\textit{Preprint} \eprint{hep-ph/0504059})

\bibitem{DiValentino:2015wba}
Di~Valentino E {\em et~al.\/} 2015  (\textit{Preprint} \eprint{1507.08665})

\bibitem{Efstathiou:2013via}
Efstathiou G 2014 {\em Mon. Not. Roy. Astron. Soc.\/} {\bf 440} 1138--1152
  (\textit{Preprint} \eprint{1311.3461})

\bibitem{Blake:2011en}
Blake C {\em et~al.\/} 2011 {\em Mon. Not. Roy. Astron. Soc.\/} {\bf 418}
  1707--1724 (\textit{Preprint} \eprint{1108.2635})

\bibitem{Beutler:2011hx}
Beutler F {\em et~al.\/} 2011 {\em Mon. Not. Roy. Astron. Soc.\/} {\bf 416}
  3017--3032 (\textit{Preprint} \eprint{1106.3366})

\bibitem{Percival:2009xn}
Percival W~J {\em et~al.\/} (SDSS) 2010 {\em Mon. Not. Roy. Astron. Soc.\/}
  {\bf 401} 2148--2168 (\textit{Preprint} \eprint{0907.1660})

\bibitem{Padmanabhan:2012hf}
Padmanabhan N {\em et~al.\/} 2012 {\em Mon. Not. Roy. Astron. Soc.\/} {\bf 427}
  2132--2145 (\textit{Preprint} \eprint{1202.0090})

\bibitem{Anderson:2013zyy}
Anderson L {\em et~al.\/} (BOSS) 2014 {\em Mon. Not. Roy. Astron. Soc.\/} {\bf
  441} 24--62 (\textit{Preprint} \eprint{1312.4877})

\bibitem{diValentino:2015inprep}
Di~Valentino E, Gariazzo S, Giusarma E and Gerbino M 2015 {\em in
  preparation\/}

\bibitem{Gariazzo:2015qea}
Gariazzo S, Lopez-Honorez L and Mena O 2015 {\em Phys. Rev.\/} {\bf D92} 063510
  (\textit{Preprint} \eprint{1506.05251})

\end{thebibliography}

\end{document}